Refractive index of dense materials

Gilbert Zalczer

SPEC, CEA, CNRS, Université Paris Saclay, CEA Saclay,  91191 Gif sur Yvette, France

We show that applying the Lorentz-Lorenz transformation to the refractive index of metals, semiconductors and insulators allows for a less empirical modeling of this refractive index.

INTRODUCTION

Optical devices are ubiquitous in our environment. They use many materials which interact with light through a complex, frequency dependent property : their refractive indices. These have been widely measured and tabulated. Several ways of empirical modeling have been proposed, usually limited to a category of materials and implying intricate formulas[1-5]. We have noticed that the "polarizability" computed by applying blindly the Lorentz-Lorenz formula can be fitted much more accurately. Indeed the imaginary part can be very well described by a few Gaussian functions and the real part computed using Kramers-Kronig relationship.

MODEL AND FITS

The dielectric constant ε of an assembly of polarizable point particle is well described by the Clausius-Mossoti formula

$$\frac{\epsilon - 1}{\epsilon + 2} = C\,\rho\alpha$$

where ρ is the density of particles α their polarizability and C a constant. Both ε and α are complex numbers and frequency dependent. In the optical domain this equation can be rewritten as

$$\frac{n^2 - 1}{n^2 + 2} = C\,\rho\alpha$$

known as Lorentz-Lorenz formula. This formula has no basis for being applied to metals, but an empirical use reveals an interesting phenomenon. The data we use originate from the book of Palik[6].

The most striking case is that of silver. The real and imaginary part of the so computed "polarizability" are shown in fig 1 as a function of the wavevector (the data are plotted in red, the fit in blue) :

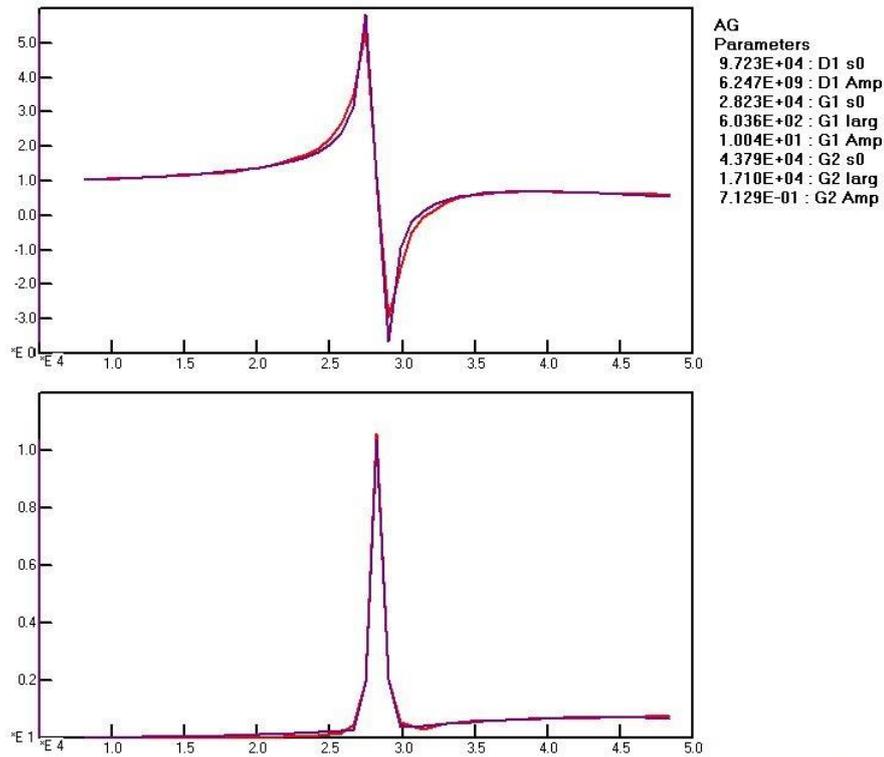

Figure 1 : Real (top) and imaginary part of the polarizabilty calculated by application of the Lorentz-Lorenz formula to the refractive index of silver as a function of the wavevector (in cm$^{-1}$)

The imaginary part is readily seen as the sum of two features : one sharp and intense near 2.8 10$^4$ cm$^{-1}$ and a broad one centered near 4.5 10$^4$ cm$^{-1}$ . These can be quite accurately fitted by Gauss curves. The fit parameters for the center wavevector (s0), the width (larg) and the amplitude (amp) of these Gauss functions (G1 and G2) are shown beside the plots. The real part cannot be deduced from the imaginary one using the Kramers-Kronig relation (the K-K transform of a Gaussian is a Dawson function)  because we have to take into account at least one other line far enough in the UV not to appear in the figure. By convention we ascribe it a width of 1 and the other parameters are reported labeled as D1. The refractive index of transparent media is due to such a line. We shall see below that the polarizability of many dense materials can be described accurately by just a small number of Gaussian shaped absorption lines. Those lying out of the observed range are seen only to the long tail of their real part. Their width cannot be determined and has been fixed to 1.

The case of gold has some similarities :

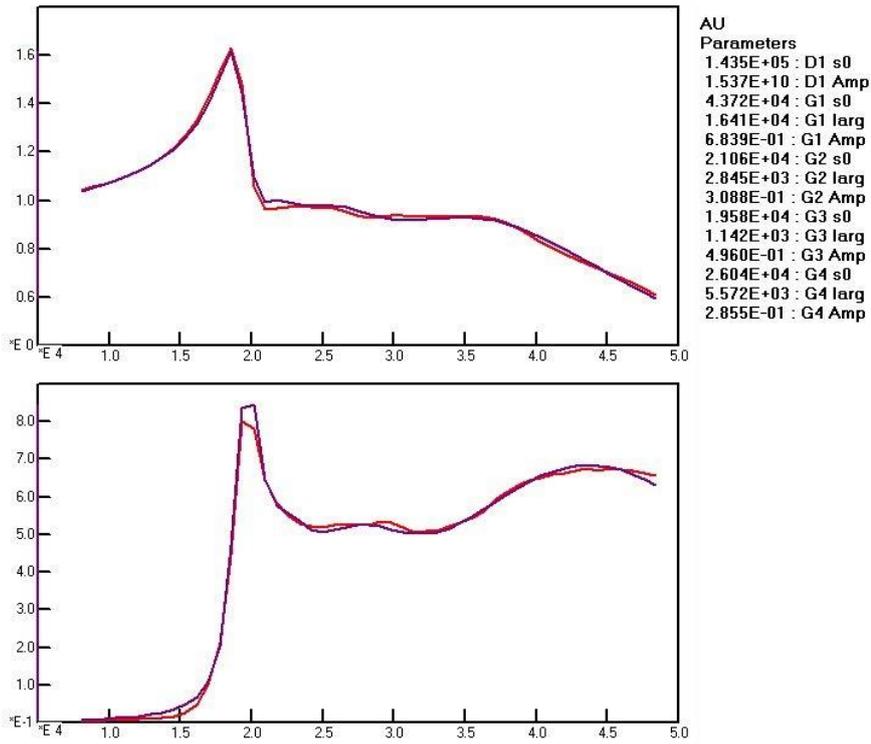

Figure 2 Polarizability of gold

A sharp peak at low wavevector and a broad one at the high side. However the sharp peak is much less intense and a wide continuum appears between them which can be accurately fitted only by introducing two additional gaussians. The same holds for copper

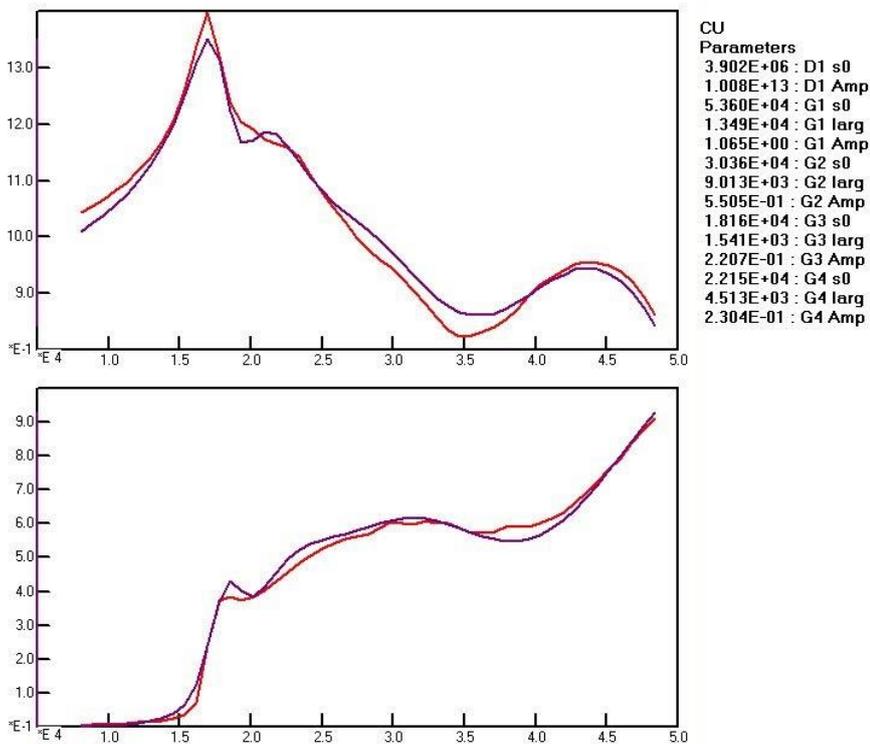

Figure 3 Polarizability of copper

The fit is a little worse. It could perhaps be improved by adding more lines but their relevance is questionable. A common feature of these three spectra is a range of zero α" at small wavevector.

Platinum and nickel can be described by one and two lines covering the visible range :

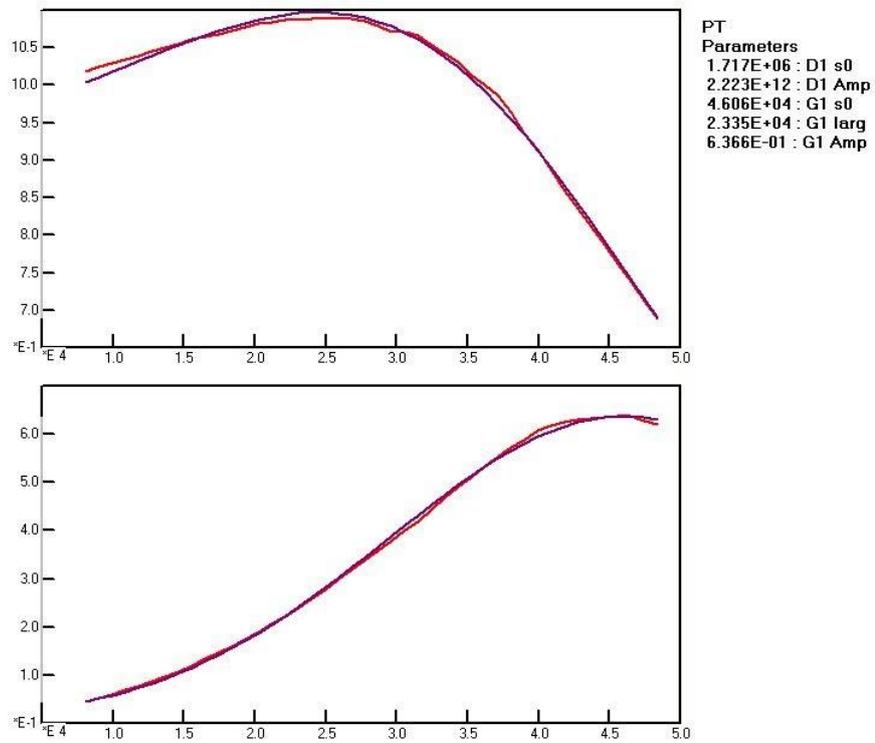

Figure 4 Polarizability of platinum

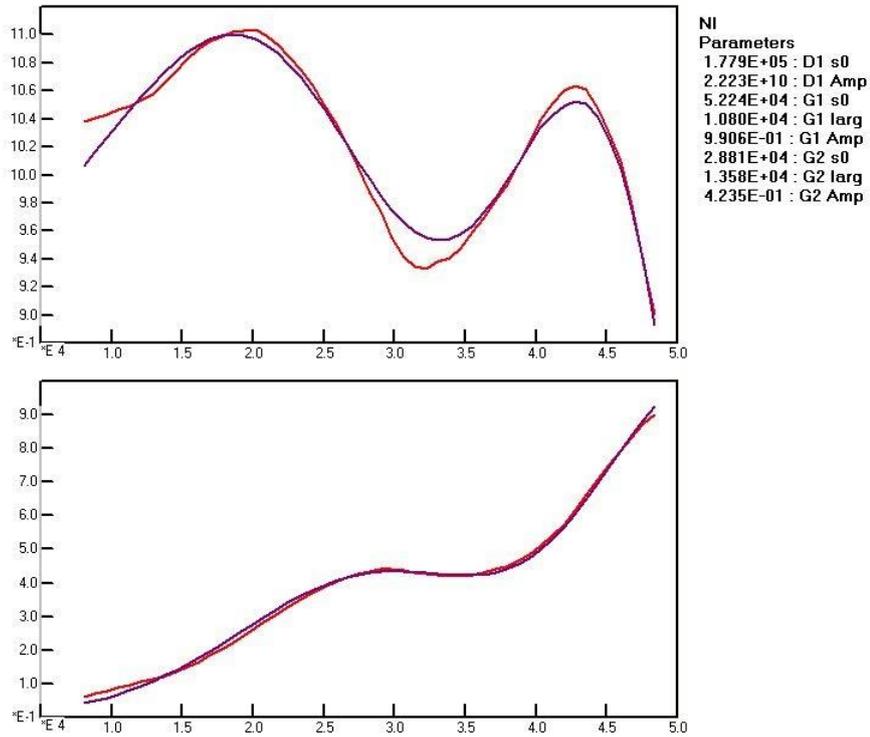

Figure 5 Polarizability of nickel

Semiconductors also can be fitted with 3 lines for silicon and 4 for germanium :

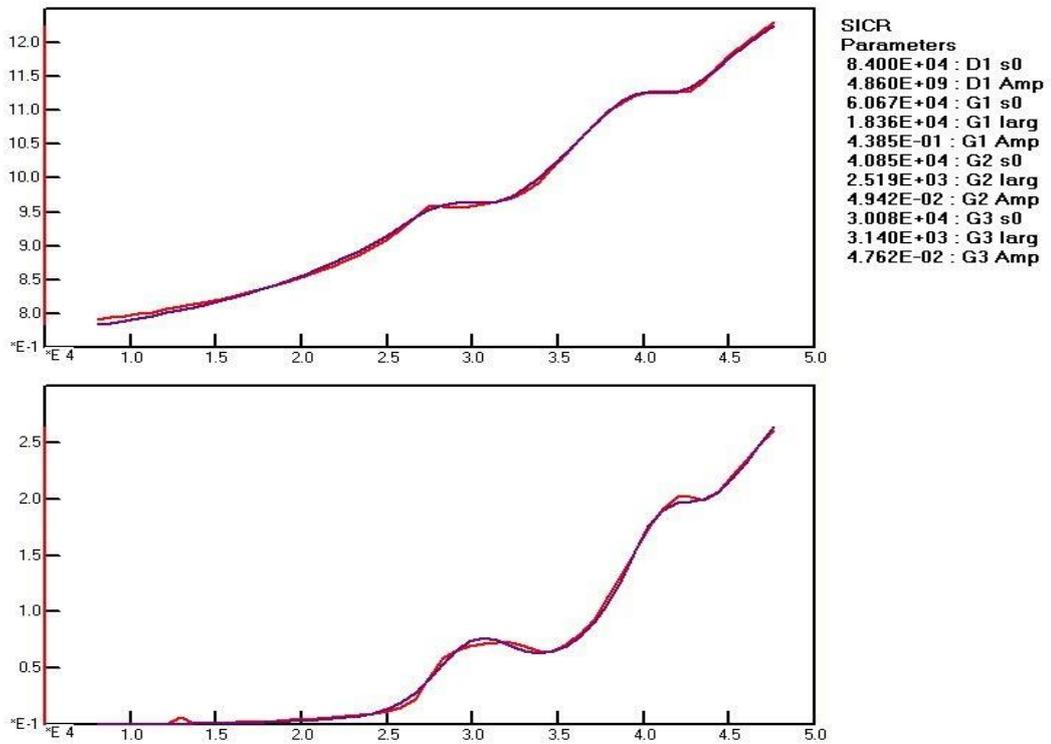

Figure 6 : Polarizability of crystalline silicon.

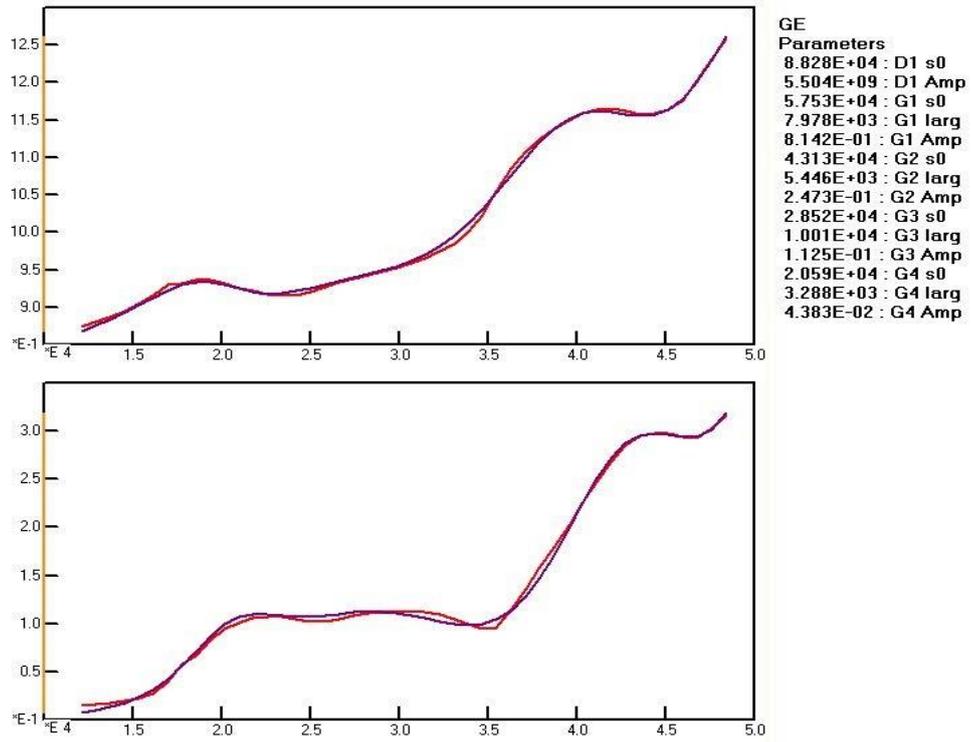

Figure 7 Polarizability of germanium

As noted above a transparent medium such as silica is due to a UV band

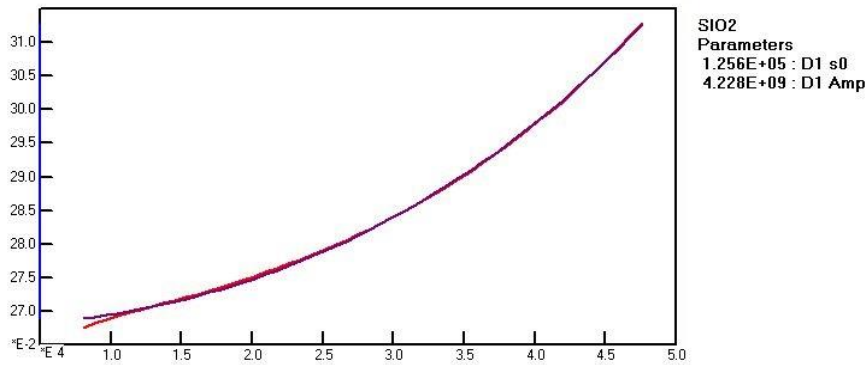

Figure 8 Polarizability of silica glass. The imaginary part is zero

While for zirconia ( ZrO2) we have also to take into account an infrared one :

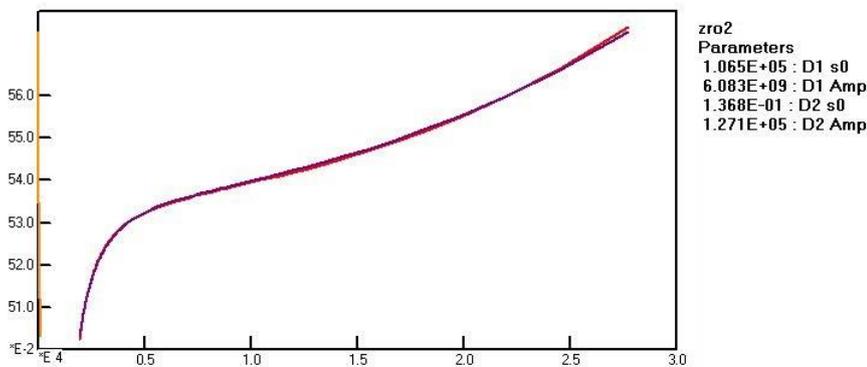

Figure 9 Polarizability of zirconia glass. The imaginary part is zero

CONCLUSION

The refractive index of dielectrics, semiconductors and metals can be interpreted and described by transforming their refractive index into a polarizability using the Lorentz-Lorenz formula. The imaginary part of this polarizability appears as a sum of a few Gaussian functions, the parameters of which can be readily fitted. Small distortions may be attributed to weaker bands but their relevance is questionable. The real part can be deduced from the Kramers-Kronig transform of the imaginary part but one has to take into account lines in the UV or in the IR, invisible in the imaginary part but who have a significant but smooth contribution in the real part. In some cases the fit for the imaginary part is good but not that of the real part. An apparent violation of the Kramers Kronig relation might signal imperfect measurements.